\documentstyle[11pt,aaspp4]{article}

\slugcomment{To appear in ApJ Letters}
\lefthead{Collins et al.}
\righthead{Evolution of X-ray Galaxy Clusters}

\def\lumin{\mbox{erg\,s$^{-1}$}}
\def\flux{\mbox{erg\,s$^{-1}$\,cm$^{-2}$}}
\def\degdeg{\mbox{deg\,$^{2}$}}
\def\perdegdeg{\mbox{deg\,$^{-2}$}}
\def\keV{\mbox{keV}}
\def\norm{\mbox{Mpc$^{-3}$\,(10$^{44}$\,\lumin)$^{\,\alpha-1}$}}
\def\fluxlimit{\mbox{$3.9\,\times\,10^{-14}\,\flux$}}
\def\passband{\mbox{$0.5 - 2.0$\,\keV}}
\def\L44{\mbox{$L_{44}$}}

\def\wurzburg{
1996, in MPE Report 263,
Proceedings of R\"ontgenstrahlung from the Universe,
ed. Zimmermann H.U., Tr\"umper J., Yorke H.
(Munich:MPE)}

\def\highz{\mbox{15}}
\def\surveyarea{17.2}
\def\surveydens{\mbox{$2.0 ^{+0.4}_{-0.3}$}}
\def\nclus{\mbox{35}}

\begin{document}

\title{On The Evolution of X-ray Clusters at High
Redshift\footnote{Based partly on data collected at the European Southern
Observatory, La Silla, Chile, and the Anglo-Australian Telescope,
Siding Spring, Australia}}

\author{C.A. Collins}
\affil{Astrophysics Group, School of Electrical Engineering, Electronics and 
Physics, Liverpool John Moores University, Byrom Street, Liverpool L3 3AF, UK}

\author{D.J. Burke\altaffilmark{2}} 
\affil{Department of Physics, University of Durham, South Road, 
Durham DH1 3LE, UK}

\author{A.K. Romer\altaffilmark{3}}
\affil{Department of Physics and Astronomy, Northwestern University,
2145 Sheridan Road, Evanston, IL 60208, USA}

\author{R.M. Sharples}
\affil{Department of Physics, University of Durham, South Road, Durham DH1 3LE, UK}

\and
\author{R.C. Nichol\altaffilmark{3}}
\affil{Department of Astronomy and Astrophysics, University of Chicago, 
5640 S. Ellis Rd., Chicago, IL 60637 USA}

\altaffiltext{2}{Present Address: Astrophysics Group, School of Electrical 
Engineering, Electronics and Physics, Liverpool John Moores University,
Byrom Street, Liverpool L3 3AF, UK}

\altaffiltext{3}{Present Address: Department of
Physics, Carnegie Mellon University, 5000 Forbes Avenue, Pittsburgh, 
PA 15213-3890, USA}

\begin{abstract}

We report on the first results from a redshift survey of 
a flux-limited sample of X-ray clusters selected serendipitously 
from the ROSAT PSPC data archive.
We spectroscopically confirm $\highz$ clusters in the range $0.3<z<0.7$, to a flux limit
of $\simeq \fluxlimit$, over a survey area of
$\surveyarea$ \degdeg.
The surface density of clusters in our survey
is $\surveydens\,\perdegdeg$, in good agreement with the
number density of cluster candidates detected using
algorithms designed to search for very extended sources.
The number of clusters detected between $0.3<z<0.7$ is consistent
with a prediction
based on a simple extrapolation of the local X-ray cluster luminosity
function, which indicates that over this redshift range no significant 
evolution in the cluster
population has taken place. These results are in conflict with 
recent claims that the number density of
X-ray clusters found in deep ROSAT PSPC pointings 
evolves rapidly beyond $z=0.3$.

\end{abstract}


\section{Introduction}

Galaxy clusters are the largest gravitationally bound structures 
in the universe
and can be observed out to high redshift, providing a rich 
source of information on the history of structure formation.  
The form of cluster evolution with look-back time is a
very sensitive probe of hierarchical structure formation models.
Assuming a CDM-like power spectrum and a self-similar
model for structure evolution, \cite{Kaiser-(86)} predicted
that the comoving number density of clusters should increase with
redshift, i.e. positive evolution.
More sophisticated analytic and N-body models,
which include gravitational instability theory and hydrodynamics,
predict a decrease in the number density of clusters with redshift 
(negative evolution) for only the very brightest clusters
(X-ray luminosities, $L_x \gtrsim 5 \times 10^{44}\,\lumin$\,),
with little evolution at lower luminosities
(e.g. \cite{Kaiser-91}; \cite{cen-ost-94}).

Until recently the only available distant cluster catalogues
were optically selected, showing little evidence for
evolution in the cluster space density out to $z \simeq 0.5$
(e.g. \cite{gunn1-86}; \cite{postman-96}).
However, one of the most intriguing results to emerge in 
recent years is the lack of high redshift
X-ray luminous clusters in 
the Einstein Extended Medium Sensitivity Survey
(EMSS, \cite{gioia-90}; \markcite{henry-(92)}Henry et al. 1992) 
and other X-ray surveys (\cite{edge-90}).
These studies concluded that there has been significant
negative evolution over short look-back times in the bright end of the cluster 
X-ray luminosity function (XLF).
The evolution found by \markcite{edge-90}Edge et al. (1990) is likely to be
the result of unfortunate sampling in the redshift range $0.1 < z < 0.15$ 
(\cite{ebeling-1995}), while the evidence for negative evolution
reported by \cite{henry-(92)} is 
critically examined by \cite{nichol-(97)} using ROSAT pointed observations of
EMSS clusters. 

We are now able to test the negative evolution scenario
using the publicly available data archive of the
ROSAT Position Sensitive Proportional Counter (PSPC).
The greater spatial resolution and sensitivity,
combined with the lower background, of the ROSAT PSPC
compared to the EINSTEIN IPC means that
cluster samples can be compiled
to a flux limit an order of magnitude
fainter than that of the EMSS.
We have carried out a survey for serendipitously detected
X-ray clusters from the PSPC data archive
in order to study the cluster population at high
redshift (\cite{Wurzburg-me}).
This project is referred to as the
Serendipitous High-redshift Archival ROSAT Cluster survey (SHARC), although 
our selection technique is completely different to that described 
in \cite{nichol-(97)}, who use a source detection algorithm based on a
wavelet transform to re-examine the EMSS cluster sample. There are several 
groups currently using the PSPC database to compile distant cluster surveys
(RIXOS, \markcite{RIXOS-(nature)}Castander et al. 1995; RDCS, \cite{rosati-95}; WARPS, 
\cite{scharf-96}). The purpose of this letter is to report on the number of
spectroscopically confirmed high redshift clusters in our survey and to 
compare our results with those of other groups.
In particular we make a direct comparison with RIXOS,
who have claimed that the dearth of high redshift clusters
in their sample strengthens the claims of negative evolution
(\markcite{RIXOS-(nature)}Castander et al. 1995).
In this letter we use
$\rm{H}_0 = 50\,\mbox{km}\,\mbox{s}^{-1}\,\mbox{Mpc}^{-1}$
and $q_0 = 0.5$.

\section{The Cluster Sample}

The ROSAT PSPC provides wide field imaging capabilities
with moderate spatial and spectral resolution
(\cite{trumper-83}).
On axis the PSPC has a PSF with a FWHM of 25\arcsec\ at
1 \keV\ and an
energy resolution $\Delta E / E \simeq 0.4$ over the range
$0.1-2.4$ \keV.
Although the instrument has a 1\arcdeg\ radius field of view,
the off axis PSF degradation and
the vignetting caused by the window support structure 
generally limit analysis to
within a radius of 18\arcmin. The X-ray emission from rich clusters of 
galaxies is spatially extended, with a surface brightness profile
well fit by a King profile:
\begin{equation}
\label{eqn:sb}
S_x(r) \propto ( 1 + (r/r_c)^2 )^{1/2-3\beta}, 
\end{equation}
with a core radius ($r_c$) of 200 -- 500 kpc and
$\beta \simeq 2/3$ (\cite{jones-84}).
Assuming no evolution in $r_c$ and $\beta$,
clusters can be resolved 
out to $z \simeq 1.0$ by the PSPC. Ideally we would like to optically 
identify all the X-ray sources in our survey, regardless of extent.
However, AGN are the primary
X-ray source population at
flux limits of $10^{-13} - 10^{-14}\,\flux$
(\cite{georgantopoulos-96}), outnumbering clusters by 10 to 1. Since
X-ray emission from AGN is unresolved by the PSPC, we can significantly
reduce the contamination these sources introduce into our survey by
restricting our attention to extended sources only.
Clusters have typical ICM temperatures of 2 -- 10 \keV\ 
(\cite{david-93}), so we can reduce the contamination from soft
sources and Galactic emission by restricting our analysis to
the \passband\ pass band.

Our survey consists of 66
deep ROSAT pointings which satisfy the following
criteria:
exposure times greater than 10 ks,
$|b| > 20\arcdeg$ and
$\delta < 20\arcdeg$.
The Galactic neutral hydrogen column density 
for these fields varies between
2 and 7 $\times\,10^{20}\,\mbox{cm}^{-2}$
which, together with the average exposure time of $18$ ks, results in an 
average flux limit $\simeq \fluxlimit$.
Due to the degradation of both the PSF and the detector
effective area with off-axis position,
we limit our survey to the central 18\arcmin\ of each field,
which results in a total survey area of
$\simeq \surveyarea$ \degdeg,
after masking out the original targets
of the observations. Briefly, the analysis begins with screening out
data taken during periods of high aspect error or particle background. 
We compute a global estimate of the background and use a
sliding box technique, with a detect cell size of 30\arcsec,
coupled with the Cash statistic (\cite{cash-79}) to search for sources.
These sources are tested for extent by comparing the photon
distribution to the PSF,
modelling both the positional and spectral dependence of the PSF.
Those sources which show significant ($\geq 3 \sigma$) angular extent
are flagged as cluster candidates. The degree
of completeness in our survey depends heavily on the extent of the
cluster X-ray emission and the sensitivity of the 
extended source algorithm (e.g. see \cite{scharf-96}).
Our detection method is designed to
pick-up moderately extended sources of size $\leq$ 1\arcmin\ which is a good 
match to the expected angular size of clusters in the range $0.3 < z < 1$. 
At this stage we do not know to what extent our completeness is 
compromised by ignoring intrinsically unresolved sources. 
However, both the surveys of Georgantopoulos et al. (1996) and RIXOS, which 
have optically identified all point sources in deep ROSAT PSPC fields,
detect very few clusters with point-like X-ray emission 
(\cite{georgantopoulos-96}; \cite{castander-priv},
private communication). In addition, we are in the process 
of carrying out simulations to estimate the completeness and 
the sensitivity of our selection method to sources of different fluxes 
with sizes ranging from point-like to very extended ($\geq$ 1\arcmin). 

We have carried out an optical identification programme of the
extended sources, obtaining both imaging and multi-object spectroscopy 
using the European Southern Observatory 3.6m and 
the Anglo-Australian 3.9m telescopes. Typically we have redshifts for $5-10$ 
galaxies per cluster. Our final sample consists of clusters which satisfy
the joint criteria of both 
extended X-ray emission and multiple redshifts, which ensures minimal 
contamination from AGN. 
Full details of the survey, including the cluster identification 
procedure, redshift reduction and an estimate of the survey completeness
from simulations will be given in a forthcoming paper 
(Burke et al., in preparation). As a result of the optical observations we 
have identified \nclus\ clusters ranging from nearby poor clusters to high 
redshift rich clusters. Figure~\ref{fig:contours} shows our four most distant 
clusters with X-ray contours overlaid on CCD R band images.

\section{Comparison with other ROSAT Surveys}

We have spectroscopically confirmed a total of \nclus\ clusters over an area of 
$\surveyarea$ \degdeg, giving a cluster surface density of 
$\surveydens\,\perdegdeg$ ($1\,\sigma$ Poisson errors taken from 
\cite{gehrels-86}).
Two surveys designed to detect very extended sources in ROSAT PSPC
pointings have recently published surface densities of extended sources
found in the \passband\  band:
RDCS use a wavelet transform algorithm and report a
surface density of $\simeq 2.5$ \perdegdeg\ at a flux limit of
$3 \times 10^{-14}$ \flux;
WARPS use a Voronoi tesselation and percolation
algorithm and detect 2.1 - 3.1 \perdegdeg\ above an intrinsic flux limit 
of $4.5 \times 10^{-14}$ \flux.
Although neither sample is fully identified
(\cite{rosati-95} report spectroscopic 
confirmation for a subset of their sample whereas
\cite{scharf-96} leave discussing their optical follow up
to a future paper),
their results are similar to our surface density
of spectroscopically confirmed clusters.

In Table~\ref{tbl:survey} we summarise our survey statistics
along with those of RIXOS and show the corresponding redshift
distributions in Figure~\ref{fig:nz}.
Despite the similarity of the surveys in terms of limiting
X-ray flux and total area surveyed, it is immediately apparent
that we detect many more high redshift
clusters than RIXOS; beyond $z = 0.3$ we find $\highz$ clusters
compared to their 4. Using Poisson statistics, the probability that this
difference could arise by chance is $2.1\times10^{-4}$ and is therefore ruled
out at a significance of $3.5\, \sigma$.
The lack of high redshift clusters in the RIXOS sample could
be due to their detection method, as suggested by \cite{scharf-96},
or due to optical misclassification.

\section{Evidence For Evolution}

We examine the redshift histogram in Figure~\ref{fig:nz}
for evidence of evolution by predicting the expected number of
clusters in the interval $0.2<z<0.7$ using the low redshift cluster XLF of 
\cite{degrandi-(96)}. For clusters at lower redshifts, the results become 
unreliable due to the small volume sampled and the uncertainties in the
XLF at low luminosities. A similar calculation was carried out by 
\cite{RIXOS-(nature)} for the RIXOS survey using the cluster XLF
of \markcite{edge-90}Edge et al. (1990).
The cluster XLF of \cite{degrandi-(96)} is well 
matched to our dataset since it is a purely X-ray selected cluster sample,
identified from the ROSAT All Sky Survey.
The differential XLF, $n(L)$, is modelled using a Schechter function,
\begin{equation}
\label{eqn:schechter}
n(\L44) = A \exp \left ( -\L44/L_{\ast} \right ) \L44^{-\alpha},
\end{equation}
where \L44\ is the X-ray luminosity in units of $10^{44}\,\lumin$,
$\alpha=1.32^{+0.21}_{-0.23}$,
$L_{\ast} = 2.63^{+0.87}_{-0.58}\,10^{44}\,\lumin$
and $A = 4.29^{+0.39}_{-0.31} \times 10^{-7}\,\norm$.

Using this model we can estimate the number of 
high redshift clusters we expect to detect
by integrating equation~\ref{eqn:schechter}
above the minimum detectable luminosity for 
each redshift interval.
The exposure time, background count rate and
a model of the off axis behaviour of the PSF (\cite{offax-psf})
are used to predict the minimum
detectable count rate at each redshift, after correcting
for the flux falling outside the 
detection cell using equation~\ref{eqn:sb},
with $\beta=2/3$ and $r_c = 250$ kpc.
We use a 6 \keV\ thermal bremsstrahlung
model to convert count rates to fluxes, which are then
corrected for Galactic absorption and k-corrected to
the \passband\ band.
Both our survey and the EMSS have a flux
limit which varies from field to field.
While this complicates the
analysis, the $V/V_{max}$ method used by \cite{henry-(92)} fully corrects
for this effect and our procedure follows the same technique. 

The results summarised in Table \ref{tbl:survey} demonstrate that in the 
redshift intervals $0.2-0.3$ and $0.3-0.7$, the number of X-ray clusters 
predicted and observed differ by only $1.3\,\sigma$ and $0.9\,\sigma$ 
respectively. Despite the possibility of residual
incompleteness our data provides no evidence for
the negative evolution reported by \cite{RIXOS-(nature)}.

To test whether uncertainties in the local XLF
can bias this analysis, we have repeated the calculation
using the cluster XLF of \cite{Wurzburg-(ebeling)},
which is not strictly X-ray selected, 
but based partly on the optical catalogues of Abell and Zwicky.
The XLF is modelled as before, with
$\alpha = 1.78 \pm 0.09$,
$L_{\ast} = (8.92 \pm 1.66)\,10^{44}\,\lumin$
and $A = (5.72 \pm 0.95) \times 10^{-7}\,\norm$, 
over the energy range $0.1-2.4$ \keV.
Converting to the \passband\ band, we predict 
$6.9^{+1.0}_{-1.6}$ clusters between $0.2<z<0.3$ and
$23.1^{+4.8}_{-4.8}$
clusters between $0.3<z<0.7$.
These results are virtually identical to those obtained above,
indicating that our predictions are insensitive to
potential biases in the local XLF.

One difference between our survey and RIXOS is that, in common with EMSS,
our flux limit varies with the exposure time and Galactic neutral
hydrogen column density of each observation,
with an rms variation of $\simeq 20 \%$,
whereas RIXOS impose 
a constant flux limit of $3\,\times\,10^{-14}\,\flux$ across all fields.
Therefore one possibility is that the `missing' clusters lie in our
deepest fields, with fluxes below the RIXOS flux limit.
However, if we replace the individual flux limits of each field
with the average value quoted in Table~\ref{tbl:survey},
the predicted number of clusters remains essentially unchanged,
$19.7^{+5.0}_{-3.9}$ compared with $19.4^{+4.9}_{-3.9}$.
Hence the deeper fields in our survey do not unduly bias our
results.

\section{Conclusions}

We have carried out a redshift survey of extended sources selected 
from the ROSAT PSPC pointed data archive over an area of
$\surveyarea$ \degdeg\ 
and to a depth of \fluxlimit\ in
the energy range \passband.
Spectroscopic follow-up observations have produced a sample of
distant X-ray selected galaxy clusters. The surface density of our 
spectroscopically confirmed cluster sample is 
$\surveydens\,\perdegdeg$, in
good agreement with the number density of cluster candidates detected using
algorithms designed to search for very extended sources (RDCS and WARPS).
In addition, the redshift histogram of our clusters
is consistent with a simple extrapolation of the local luminosity 
function of X-ray clusters from \cite{degrandi-(96)}.
This indicates that no substantial negative evolution has taken place in
the cluster population over the redshift range $0.3<z<0.7$.

After this manuscript was accepted we spectroscopically confirmed a further
cluster at $z=0.672$ from service observations carried out on the 4.2m WHT.
This completes the identification of extended sources and brings the
number of clusters in our survey which are beyond $z=0.3$ to 16, 
which marginally increases the discrepancy between our sample and
that of RIXOS.
 
\acknowledgements

This research has made use of data obtained from the Leicester Database
and Archive Service at the Department of Physics and Astronomy,
Leicester University, UK. We thank Sabrina De Grandi for providing
us with the errors on the ESO KP cluster XLF. Mel Ulmer and Brad
Holden are acknowledged for their support of the SHARC
survey. CAC and DJB acknowledge PPARC for an Advanced Fellowship and
Postgraduate Studentship respectively. AKR acknowledges support from
NASA ADP grant NAG5-2432. We thank the staff at both the European
Southern Observatory and the Anglo-Australian Observatory for use of
these facilities.

\clearpage

\figcaption[]{
X-ray contours, for the \passband\ band 
overlaid on CCD R band exposures
of our 4 most distant clusters. The images 
are 10 minute exposures taken at the ESO 
3.6m. Axis units are in arcminutes and
the X-ray images have been lightly smoothed
and begin at $3 \sigma$ above the background
with a step size of $3 \sigma$.
A is at $z = 0.576$, B at $z = 0.550$, C at
$z = 0.505$ and D at $z = 0.484$.
At these redshifts 1 arcminute subtends
$\sim 440$ kpc.
\label{fig:contours}}

\figcaption[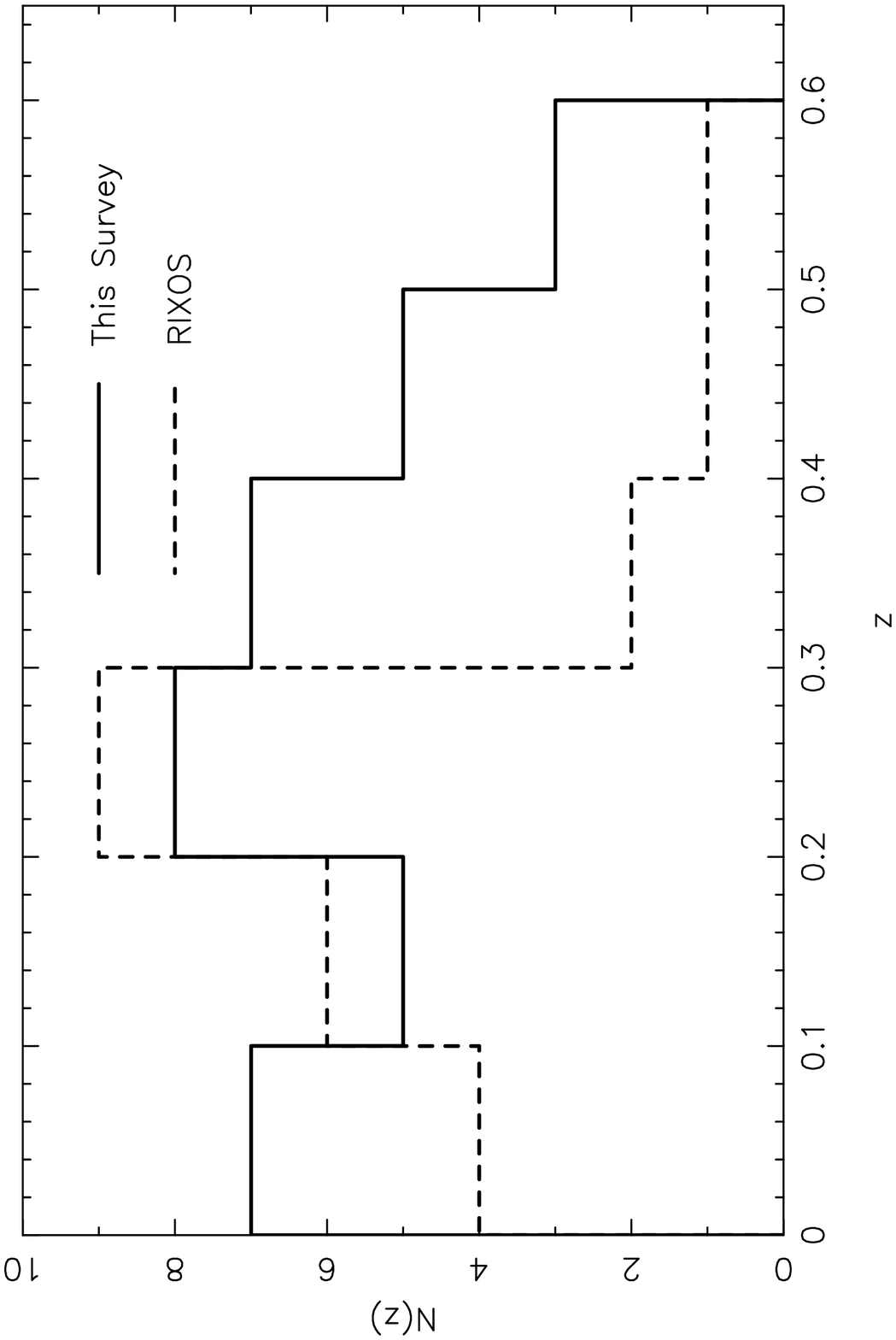]{
Comparison of our complete cluster redshift distribution
(solid line)
with that of the RIXOS survey (dotted line).
The RIXOS data are taken from
\protect \cite{RIXOS-(nature)}
for $z \geq 0.2$ and
\protect \markcite{castander-thesis}Castander (1996b) for $z < 0.2$.
\label{fig:nz}}

\clearpage

\begin{table*}
\begin{center}
\begin{tabular}{ccccc}
Survey & $\Omega$  & Flux limit & $N_c$ & $N_c$ \\
       & ($\degdeg$) & ($10^{-14}\,\flux$) & 
($0.2<z<0.3$) & ($0.3<z<0.7$) \\
\tableline
This Work & \surveyarea & 3.9 & $8 ^{+4.0}_{-2.8}$ ($4.3^{+1.6}_{-1.2}$) &
$\highz^{+5.0}_{-3.8}$ ($19.4^{+4.9}_{-3.9}$) \\
RIXOS     & 14.9        & 3.0 & $9^{+4.1}_{-2.9}$ (8) & $4^{+3.2}_{-1.9}$ 
(25) \\
\end{tabular}
\end{center}

\caption{
Comparison between this work and RIXOS for all $z > 0.2$ clusters.
The errors on the number of clusters found are Poisson.
The numbers in brackets for our survey are the predictions for the number of 
clusters
observed based on a model of our selection function and 
the XLF of \protect \cite{degrandi-(96)},
along with the $1 \sigma$ errors from the simulations.
The values for RIXOS have been taken from
\protect \cite{RIXOS-(nature)},
which uses the XLF of \protect \markcite{edge-90}Edge et al. (1990).}
\label{tbl:survey}
 \end{table*}


\clearpage

\begin{thebibliography}{}
\bibitem [Burke et al. 1996]{Wurzburg-me}
Burke, D. J., Collins, C. A., Nichol, R. C., Romer, A. K.,
Holden, B., Sharples, R. M., \& Ulmer, M. P.
\wurzburg, 569
\bibitem [Cash 1979]{cash-79}
Cash, W. 1979, \apj, 228, 939
\bibitem [Castander et al. (1995)]{RIXOS-(nature)}
Castander, F. J., et al. 1995, \nat, 377, 39
\bibitem [Castander 1996a]{castander-priv}
Castander, F. J. 1996a, private communication
\bibitem [Castander 1996b]{castander-thesis}
Castander, F. J. 1996b, Ph.D. thesis, Cambridge University
\bibitem [Cen \& Ostriker 1994]{cen-ost-94}
Cen, R., \& Ostriker, J. P. 1994, \apj, 429, 4
\bibitem [David et al. 1993]{david-93}
David, L. P., Slyz, A., Jones, C., Forman, W., Vrtilek, S. D.,
\& Arnaud, K. A. 1993, \apj, 412, 479
\bibitem [De Grandi (1996)]{degrandi-(96)}
De Grandi, S. \wurzburg, 577
\bibitem [Ebeling et al. 1995]{ebeling-1995}
Ebeling, H., B\"{o}hringer, H., Briel, U. G., Voges, W., Edge, A. C., 
Fabian, A. C., Allen, S. W., \& Huchra, J. P. 1995, in Wide Field Spectroscopy 
\& The Distant Universe, ed. Maddox, S. J., Arag\'{o}n-Salamanca, A.
(World Scientific: Singapore), 221
\bibitem [Ebeling et al. (1996)]{Wurzburg-(ebeling)}
Ebeling, H., Allen S. W., Crawford, C. S., Edge, A. C., 
Fabian, A. C., B\"ohringer, H., Voges, W., \& Huchra, J. P. 
\wurzburg, 579
\bibitem [Edge et al. 1990]{edge-90}
Edge, A. C., Stewart, G. C., Fabian, A. C., \& Arnaud, K. A.
1990, \mnras, 245, 559
\bibitem [Gehrels 1986]{gehrels-86}
Gehrels, N. 1986, \apj, 303, 336
\bibitem [Georgantopoulos et al. 1996]{georgantopoulos-96}
Georgantopoulos, I., Stewart, G. C., Shanks, T., Boyle, B. J.,
\& Griffiths, R. E. 1996, \mnras, 280, 276
\bibitem [Gioia et al. 1990]{gioia-90}
Gioia, I. M., Henry, J. P., Maccacaro, T., Morris, S. L., Stocke,
J. T., \& Wolter, A. 1990, \apjl, 356, L35
\bibitem [Gunn, Hoessel \& Oke 1986]{gunn1-86}
Gunn, J. E., Hoessel, J., \& Oke, J. B. 1986, \apj, 306, 30
\bibitem [Hasinger et al. 1994]{offax-psf}
Hasinger, G., Boese, G., Predehl, P., Turner, J. T., Yusaf, R.,
Georgem I. M., \& Rohrbach, G.
1994, MPE/OGIP Calibration Memo, CAL/ROS/93-015
\bibitem [Henry et al. (1992)]{henry-(92)}
Henry, J. P., Gioia, I. M., Maccacaro, T., Morris, S. L., Stocke, J. T., 
\& Wolter, A. 1992, \apj, 386, 408
\bibitem [Jones \& Forman 1984]{jones-84}
Jones, C., \& Forman, W. 1984, \apj, 276, 38
\bibitem [Kaiser (1986)]{Kaiser-(86)}
Kaiser, N. 1986, \mnras, 222, 323
\bibitem [Kaiser 1991]{Kaiser-91}
Kaiser, N. 1991, \apj, 383, 104
\bibitem[Nichol et al. (1997)]{nichol-(97)}
Nichol, R. C., Holden, B. P., Romer, A. K., Ulmer, M. P.,
Burke, D. J., \& Collins, C. A. 1997, \apj, in press (astro-ph/9611182)
\bibitem [Postman et al. 1996]{postman-96}
Postman, M., Lubin, L. M., Gunn, F. E., Oke, J. B.,
Hoessel, J. G., Schneider, D. P., \& Christensen, J. A.
1996, \aj, 111, 615
\bibitem [Rosati et al. 1995]{rosati-95}
Rosati, P., Della Ceca, R., Burg, R.,
Norman, C., \& Giacconi, R. 1995, \apjl, 445, L11
\bibitem [Scharf et al. 1996]{scharf-96}
Scharf, C. A., Jones, L. R., Ebeling, H., Perlman, E., Malkan, M., \& Wegner, 
G. 1996, \apj, in press
\bibitem [Tr\"umper 1983]{trumper-83}
Tr\"umper, J. 1983, Adv. Space Res., 2(4), 241
\end{thebibliography}
\end{document}